\documentclass[final,5p,times]{elsarticle}
\usepackage{amssymb}
\usepackage{epsfig}
\usepackage{graphicx}
\usepackage{longtable}
\journal{Nuclear Instruments and Methods in Physics Research, Section A}
\begin{document}
\begin{frontmatter}
%
%
%
\title{The flaring blazars of the first 1.5 years of the AGILE mission}
%
%
%
%
%
%
\author[label1]{Pacciani L.\corref{cor1}}
\cortext[cor1]{corresponding author}
\ead{luigi.pacciani@iasf-roma.inaf.it}
\author[label5]{Bulgarelli A.}
\author[label3,label4]{Chen A.W.}
\author[label1,label2]{D'Ammando F.}
\author[label1]{Donnarumma I.}
\author[label3]{Giuliani A.}
\author[label6]{Longo F.}
\author[label13]{Pucella G.}
\author[label1,label2]{Tavani M.}
\author[label22]{Vercellone S.}
\author[label1,label2]{Vittorini V.}
%
%
%
%
%
%
\author[label1]{Argan A.}\author[label6]{Barbiellini G.}
\author[label7]{Boffelli F.}
\author[label3]{Caraveo P.}
\author[label7]{Cattaneo P.W.}
\author[label1]{Cocco V.}
\author[label1]{Costa E.}
\author[label1]{De Paris G.}\author[label1]{Del Monte E.}
\author[label5]{Di Cocco G.}
\author[label1]{Evangelista Y.}\author[label4,label18]{Ferrari A.}
\author[label1]{Feroci M.}\author[label3]{Fiorini M.}
\author[label2,label4]{Froysland T.}\author[label5]{Fuschino F.}
\author[label8]{Galli M.}\author[label5]{Gianotti F.}
\author[label5]{Labanti C.}\author[label1]{Lapshov I.}
\author[label1]{Lazzarotto F.}
\author[label9]{Lipari P.}
\author[label5]{Marisaldi M.}\author[label10]{Mastropietro M.}
\author[label3]{Mereghetti S.}
\author[label5]{Morelli E.}\author[label6]{Moretti E.}
\author[label11]{Morselli A.}
\author[label20]{Pellizzoni A.}\author[label3]{Perotti F.}
\author[label1,label2,label11]{Piano G.}\author[label2,label11]{Picozza P.}
\author[label21]{Pilia M.}\author[label1]{Porrovecchio G.}
\author[label12]{Prest M.}
\author[label13]{Rapisarda M.}\author[label7]{Rappoldi A.}
\author[label1]{Rubini A.}
\author[label1,label2]{Sabatini S.}
\author[label1]{Soffitta P.}
\author[label5]{Trifoglio M.}\author[label1]{Trois A.}
\author[label6]{Vallazza E.}
\author[label3]{Zambra A.}\author[label9]{Zanello D.}
\author[label19]{Antonelli L.A.}
\author[label14]{Colafrancesco S.}
\author[label14]{Giommi P.}
\author[label14]{Pittori C.}
\author[label14]{Verrecchia F.}
\author[label14]{Santolamazza P.}
\author[label15]{Salotti L.}
\address[label1] {INAF/IASF-Roma, I-00133 Roma, Italy}\address[label2] 
{Dip. di Fisica, Univ. Tor Vergata, I-00133
Roma,Italy} \address[label3] {INAF/IASF-Milano, I-20133 Milano, Italy}
\address[label4] {CIFS-Torino, I-10133 Torino, Italy}
\address[label5] {INAF/IASF-Bologna, I-40129 Bologna, Italy}
\address[label6] {Dip. Fisica and INFN Trieste, I-34127 Trieste,
Italy} \address[label7] {INFN-Pavia, I-27100 Pavia, Italy}
\address[label8] {ENEA-Bologna, I-40129 Bologna, Italy}
\address[label9] {INFN-Roma La Sapienza, I-00185 Roma, Italy}
\address[label10] {CNR-IMIP, Roma, Italy} \address[label11] {INFN
Roma Tor Vergata, I-00133 Roma, Italy} \address[label12] {Dip. di
Fisica, Univ. Dell'Insubria, I-22100 Como, Italy}
\address[label13] {ENEA Frascati,  I-00044 Frascati (Roma), Italy}
\address[label14] {ASI Science Data Center, I-00044
Frascati(Roma), Italy} \address[label15] {Agenzia Spaziale
Italiana, I-00198 Roma, Italy} \address[label16] {Osservatorio
Astronomico di Trieste, Trieste, Italy} 
\address[label18]
{Dip. Fisica, Universit\'a di Torino, Turin, Italy}
\address[label19] {INAF-Osservatorio Astron. di Roma, Monte Porzio
Catone, Italy} \address[label20] {INAF-Osservatorio Astronomico di
Cagliari, localita' Poggio dei Pini, strada 54, I-09012 Capoterra,
Italy} \address[label21] {Dipartimento di Fisica, Universit\'a
dell'Insubria, Via Valleggio 11, I-22100 Como, Italy}
 \address[label22] {INAF-IASF Palermo, Via Ugo La Malfa 153, I-90146 
Palermo, Italy}
%
%
%
\begin{abstract}
We report the AGILE $\gamma$-ray observations and the results of the multiwavelenght campaigns
on seven flaring blazars detected by the mission:
During two multiwavelenght campaigns, we observed $\gamma$-ray activity
from two Flat Spectrum Radio Quasars of the Virgo region, e.g. 3C 279 and 3C 273 (the latter being the first
extragalactic source simultaneously observed with the $\gamma$-ray telescope and
the hard X ray imager of the mission).
Due to the large FOV of the AGILE/GRID instrument, we achieved an almost
continuous coverage of the FSRQ 3C 454.3. The source showed
flux above 10$^{-6}$ photons/cm$^2$/s (E $>$ 100 MeV) and showed day by day
variability during all the AGILE observing periods. In the EGRET era, the
source was found in high $\gamma$-ray activity only once.
An other blazar, PKS 1510-089 was frequently found in high $\gamma$-ray activity.
S5 0716+71, an intermediate BL Lac object, exhibited a very high $\gamma$-ray
activity and fast $\gamma$-ray variability during a period of intense optical activity.
We observed high $\gamma$-ray activity from W Comae, a BL Lac object, and
Mrk 421, an high energy peaked BL Lac object. For this source, a
multiwavelenght campaign from optical to TeV has been performed.

\end{abstract}


\end{frontmatter}

%
\section*{Introduction}
AGILE \cite{tavanagile} is a small satellite mission for $\gamma$-ray Astrophysics;
the payload houses a pair conversion $\gamma$-ray telescope (GRID, \cite{grid}) covering the range 30 MeV - 50 GeV
with a field of view of $\sim $ 2.5 sr,
an hard X-ray monitor (SuperAGILE \cite{superagile}) covering the range 18 keV - 60 keV with a field of view of $\sim$ 0.8 sr,
a mini-calorimeter (MCAL \cite{mcal}) covering the range 300 keV - 100 MeV, and an Anticoincidence System (AC).
AGILE is observing the $\gamma$-ray sky since its launch on April 2007.\\
Among the extragalactic objects, the GRID telescope is particularly suited to the study of
Blazars. They are Active Galactic Nuclei, with jet orientation close to the line of sight
of the observer \cite{urry}.
These objects emit from radio band to $\gamma$-ray (for 3C 279, emission has been detected in the TeV range \cite{279tev}).
Blazars show high $\gamma$-ray activity and variability above 100 MeV.
The source of the emission is believed to be the non-thermal emission from the relativistic jet, overwhelming the
thermal components,
but the emission mechanisms are still debated. Hadronic and Leptonic models were developed to explain the observations.
In the framework of leptonic model, the $\gamma$-ray emission is explained with the Inverse-Compton mechanism.
Two scenarios are invoked : The Synchrotron Self Compton (SSC) of the relativistic electrons from the jet and the photon field
(coming from the synchrotron radiation produced by the same relativistic
electron population itself, see \cite{maraschi1992}, \cite{marscher1992}) . The other scenario, the External Compton (EC), foresees an external photon field as the seed for the
inverse-Compton process by the relativistic electrons. The external photon field can be identified
with the thermal radiation from the disk at least reprocessed by the Broad Line Regions (BLR)
or by the jet itself, or with the thermal radiation from an hot corona (see , \cite{sikora1994}).\\
To understand the emission mechanisms of Blazars, wide band observations (from radio to $\gamma$-ray observations),
and long term coverage are needed.
The AGILE/GRID telescope with its wide field of view, permits long term observation of the sources.
The GRID has detected several known Blazars in flare. To assure wide energy band coverage,
multiwavelenght campaigns have been organized by the AGILE/AGN working group both as scheduled multiwalenght campaigns and
as ToO for the sources detected in flare (with multiband observations triggered in optical, in Gamma, or in X rays).
\section{The Flaring Blazar sample}
\begin{table*}
  \caption{List of AGILE flaring Blazars. $^*$This $\gamma$-ray source is a Blazar candidate, with SDSS J123932.75+044305 as possible optical counterpart.}
\label{tab:sources_list}
\centering
\begin{tabular}{l | c | c | c  }
\hline\hline
Name & Period              &  S/NR  &  ATel  \\
     & \emph{start : stop} &        &        \\ \hline
%
%
S5 0716+714  & 2007-09-04 : 2007-09-23 &  9.6 &  1221      \\
             & 2007-10-24 : 2007-11-01 &  6.0 &            \\ \hline
MRK 0421     & 2008-06-09 : 2008-06-15 &  4.5 & 1574, 1583 \\ \hline
W Comae      & 2008-06-09 : 2008-06-15 &  4.0 & 1582       \\ \hline
PKS 1510-089 & 2007-08-23 : 2007-09-01 &  5.6 & 1199       \\
             & 2008-03-01 : 2008-03-20 &  7.0 & 1436       \\ \hline
3C 273       & 2007-12-16 : 2008-01-08 &  4.6 &            \\ \hline
3C 279       & 2007-07-09 : 2007-07-13 & 11.1 &            \\
             & 2007-12-16 : 2008-01-08 &  4.1 &            \\ \hline
3EG J1236+0457$^*$    & 2007-12-16 : 2008-01-08 &  5.2 &            \\ \hline
3C 454.3     & 2007-07-24 : 2007-07-30 & 13.8 & 1160, 1167 \\
             & 2007-11-10 : 2007-12-01 & 19.0 & 1278, 1300 \\
             & 2007-12-01 : 2007-12-16 & 21.3 &            \\
             & 2008-05-10 : 2008-06-30 & 15.0 & 1545, 1581, 1592 \\
             & 2008-07-25 : 2008-08-15 & 12.1 & 1634       \\ \hline
\end{tabular}
\end{table*}
During the first 1.5 years of observations of AGILE, the main part of the
observing plan was reserved to galactic sources, and particularly to the Cygnus field.
The satellite was pointed outside the galactic plane only few times,
thence the sample is mainly constituted by sources with position not to far from the galactic plane.
Nevertheless, due to the AGILE response to ToO requests (from the GASP consortium, or from VERITAS \cite{veritas}),
A few sources have been detected far from the galactic plane.
All the optical observatories are located in the northern hemisphere, thence the sources are mainly
located in the northern sky, or near the equatorial plane.
The list of flaring Blazars detected with AGILE is reported in table \ref{tab:sources_list}, together with the observing periods.
%
%
%
\section{3C 454.3}
\begin{figure}
\centering
\begin{tabular}{c}
\psfig{figure=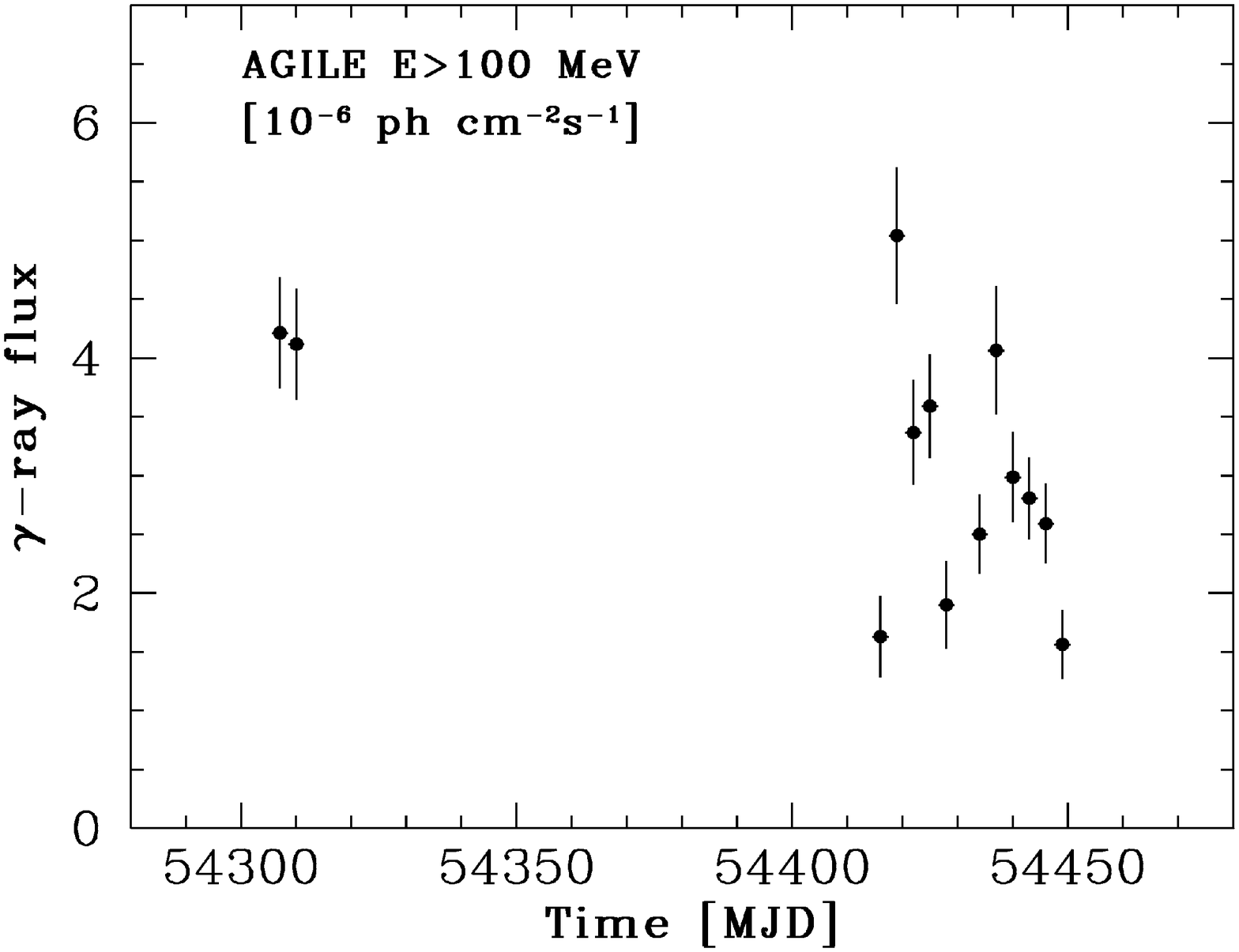,width=8.65cm,height=3.cm} \\ \psfig{figure=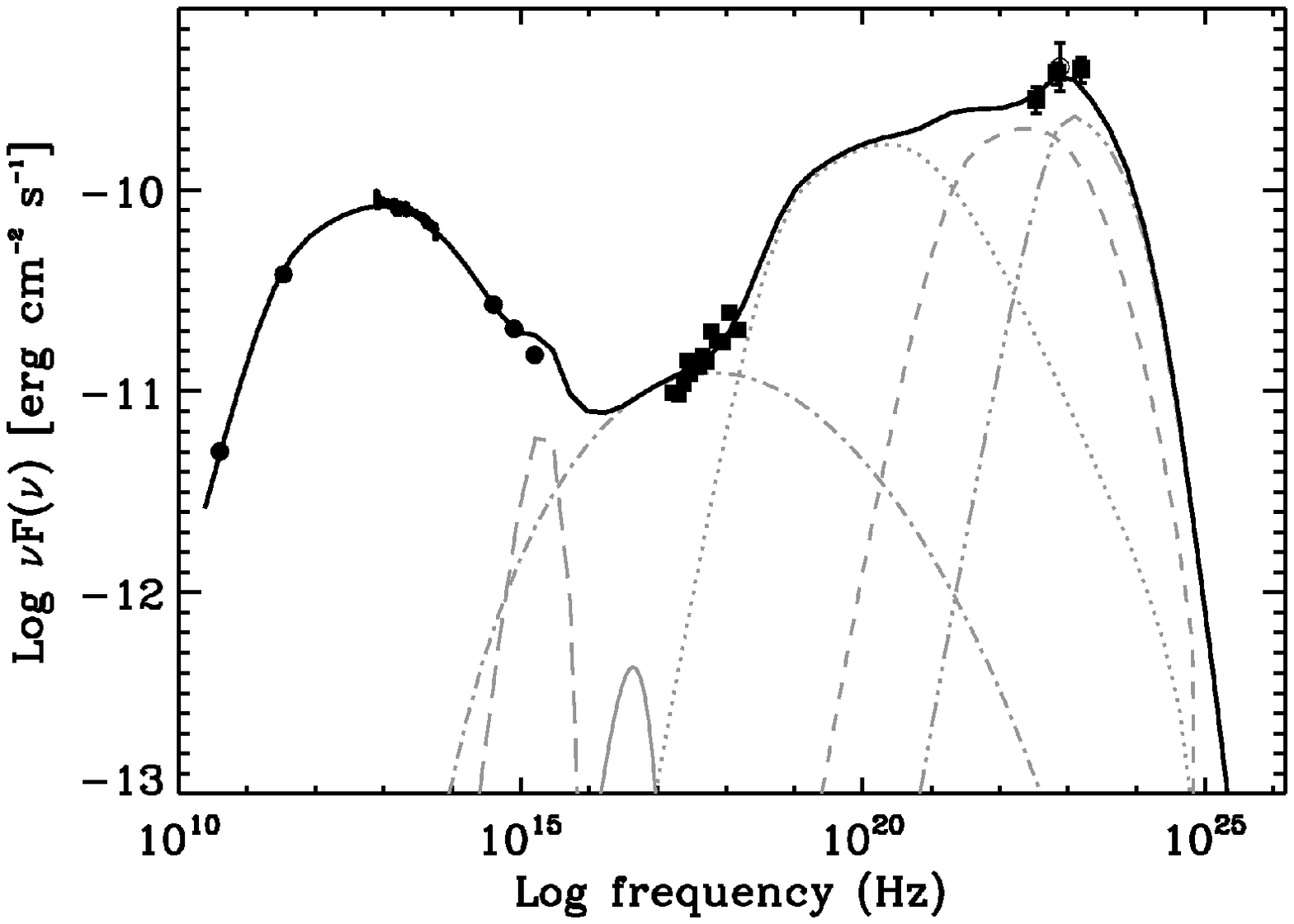,width=8.cm,height=3.cm}
\end{tabular}
\caption{
 Top panel:  Light Curve of 3C 454.3 for the period July - Dec 2007. 
 Bottom panel: Spectral Energy Distribution of 2007 December the 13$^{th}$
}
\label{fig:3c454}
\end{figure}

3C 454.3 is the flaring blazar that AGILE observed for the longest time.
This Flat Spectrum Radio Quasar exhibited $\gamma$-ray activity during all the AGILE
observations, reachining fluxes of 400 - 500 ph/cm$^2$/s (E$>$100 MeV), see \cite{porcellone454l}, \cite{porcellone454}, and \cite{donnarumma454}.
In the EGRET era, the source was detected in high $\gamma$-ray actvity only once.
The first AGILE observation of the source was in July 2007. The satellite was pointed to the
source according to a Target of Opportunity triggered by an extremely bright optical flare.
In November and December 2007, we still found the source in high state and we organized
more than one multiwavelenght campaign. \emph{Swift}, INTEGRAL and \emph{Suzaku} satellites, as well as
the REM telescope and the WEBT consortium were involved in the campaigns.
The $\gamma$-ray flux appears variable on 12-24 h time scale, but the poor statistics prevents us
to integrate on shorter time scales, and sub-daily variability can not be excluded.
The optical flux appears extremely variable, with rise times as fast as few hours (for flux increase of
tenths of magnitudes).
The $\gamma$-ray light curve for the period July - December 2007 is reported in the top panel of fig \ref{fig:3c454}.
The complete dataset for the first 1.5 years of observations will be reported in a forthcoming paper \cite{porcellone454_1.5}. 
The time correlations analysis between optical (R-band) and $\gamma$-ray, for the 1.5 year of observations, shows a delay
of 0.5-1 day of $\gamma$-ray respect to the optical.
%
All the Spectral Energy Distribution we modelled exhibited the predominant role of the External Compton component with seed photons reprocessed by the Broad Line Regions
A similar high energy contribution was found, for example, in the 90's multiwavelengths campaigns on the FSRQ 3C 279 \cite{hartman2001}.  
Interestingly, in a major flare (2007 December the 13$^{th}$), this high energy component of the SED model
was not enough to explain the data. the Spectral Energy Distribution achieved with simultaneous data is reported
in the bottom panel of fig \ref{fig:3c454} togheter with the adopted model.
To reproduce the broad band features, we were forced to add a further high energy component in the model,
describing the External Compton from an hot corona.
\section{The Blazars of the Virgo Region}
AGILE pointed the Virgo Region twice in the first 1.5 years of observations.
The first observation was performed during the Science Validation Phase of the experiment, in response to a ToO triggered by the detection of an optical flare of 3C 279.
We observed 3C 279 in an intermediate spectral state ($\Gamma \sim 2.2$, see \cite{giuliani2009}) with respect to the campaigns reported by Hartman \cite{hartman2001}.
In all that campaigns (but one, the P9 observation) the High Energy contribution of the External
Compton (with seed photons reprocessed by the Broad Line Regions) is used. Only in the last EGRET campaign (P9) the source was observed with a soft $\gamma$-ray spectra, thence with a negligible contribution from
External Compton with radiation reprocessed by the BLR.\\
The second AGILE observation of the field lasted for 3 weeks, starting from the 2nd half of December 2007. We detected two FSRQs:
3C 273 \cite{pacciani2009} and 3C 279, and an unidentified EGRET source: 3EGJ1236+0457. Simultaneous data were collected with REM, RXTE, INTEGRAL, and \emph{Swift} for the two FSRQs.
We observed $\gamma$-ray variability for all the sources.  We found 3C 273 very bright in X-ray, allowing us to disentangle the Seyfert-like contribution from the jet emission.\\
\begin{figure}
\psfig{figure=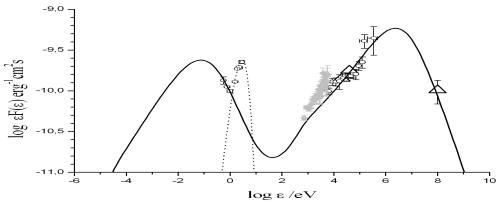,width=8.cm,height=3.cm}
\caption{
  Spectral Energy Distribution of 3C 273 for the second week of observation. Triangles are for AGILE data.
  The grey data refers to the XRT observations, performed in the third week.
  The line is the model. Where not visible the energy range is smaller than the symbol.
}
\label{fig:3c273_sed}
\end{figure}
The SED of 3C 273 is reported in figure \ref{fig:3c273_sed}, for the second week of observation of the source (when the source was brighter in $\gamma$-ray). The flux in each energy
band is similar (within 20\%) to the one measured in the multiwavelenght campaign of June 1991 \cite{licti1995}, when $\gamma$-ray variability was not observed. The observed source multiband variability can be explained as a deceleration
episode of the jet, as foreseen in \cite{sikora2001}.\\
We detected 3EGJ1236+0457, an unidentified $\gamma$-ray source, at a flux almost an order of magnitude higher than the mission averaged flux reported in the third EGRET catalogue.
The deep exposure of INTEGRAL allows us to fix deep upper-limit for the hard X flux of this source. Recently the $\gamma$-ray source was associated with an  AGN of the Sloan Digital Survey
(SDSS J123932.75+044305 located at z=1.76, see Astronomical Telegram 1888, 1892). A faint source appears in the INTEGRAL/OMC data (V band) at the coordinates of SDSS J123932.75+044305.\\
We didn't detect the blazar 4C 04.42, obtaining for the 3 weeks of observation an upper limit comparable to the EGRET one, evaluated from the integration over all the EGRET dataset.
\section{PKS 1510-089}
\begin{figure}
\centering
\begin{tabular}{c}
\psfig{figure=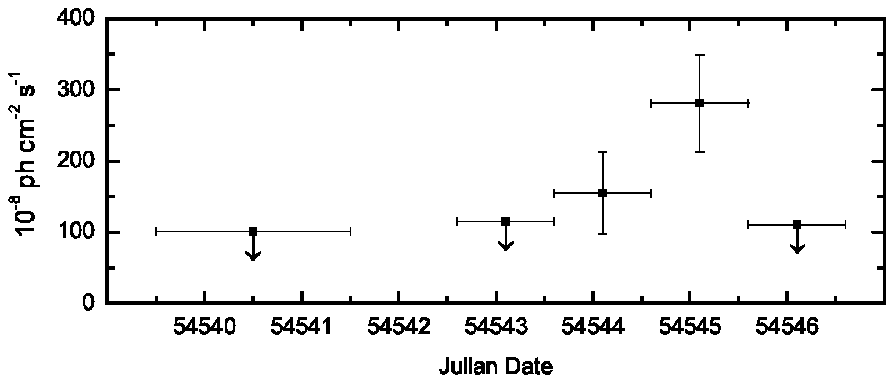,width=8.25cm,height=3.cm} \\ \psfig{figure=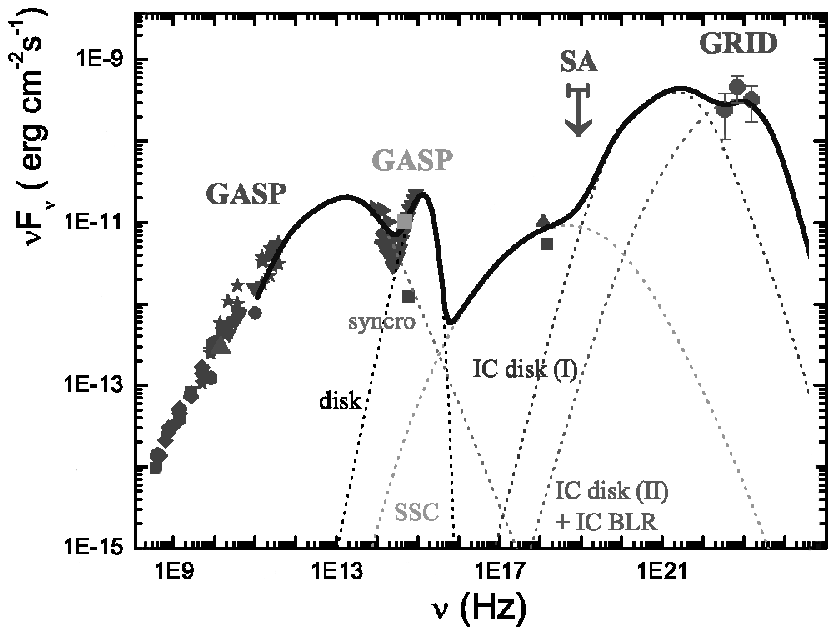,width=9.2cm,height=3.cm}
\end{tabular}
\caption{
  Top panel: Light Curve of PKS 1510-089 in the second week of March 2008. 
  Bottom panel: Spectral Energy Distribution of PKS 1510-089 of August 2007. 
}
\label{fig:pks1510}
\end{figure}
PKS 1510-089 is a Flat Spectrum Radio Quasar at z=0.361, never detected in
high $\gamma$-ray activity during the EGRET era. The measured flux was at a
level of (13-49) $\cdot 10^{-8}$ photons/cm$^2$/s (E $>$ 100 MeV)
\cite{egret3}. The $\gamma$-ray spectra was always soft, with photon index $\Gamma = (2.47 \pm 0.21)$. 
The source seems to show several emission features, as the Little Blue Bump, the Big Blue Bump, the Soft Excess, in addition to the  typical jet emission features of a blazar.
This source was detected in high $\gamma$-ray activity for the first time by AGILE in August 2007\cite{pucella}, and  underwent intense flaring activity during the AGILE observations \cite{dammando1510},
reaching a flux of 500$\cdot 10^8$ ph/cm$^2$/s (E $>$ 100 MeV) in March 2009 \cite{dammando1510bis}.
As an example, we plot the light curve for 1 week of AGILE observation in the top panel of figure \ref{fig:pks1510}. The complete data set will be shown in \cite{dammando1510}.
%
%
The $\gamma$-ray spectra during AGILE observations is harder then previous measurements. For example, in 2007, we obtained a photon index $\Gamma = (1.98 \pm 0.27)$ \cite{pucella}.
We obtained simultaneous data in radio and optical from the GASP-WEBT consortium, and in X-ray with \emph{Swift}. The simultaneous Spectral Energy Distribution obtained
in 2007 August campaign is reported in the bottom panel of figure \ref{fig:pks1510}.
The high flux in $\gamma$-ray can not be accounted for in the model without  including a further component in the $\gamma$-ray range.
Similarly to 3C 454.3 and 3C 279, we associated this component with  External Compton  with seed photons reprocessed by the Broad Line Regions
(assuming 10\% of reprocessed radiation continuum). Figure \ref{fig:pks1510} (bottom panel) shows the model components we added to fit the data.
%
%
\section{S5 0716+714}
S5 0716+714 is an intermediate BL Lac object, at z $\sim$ 0.3. AGILE detected the source twice: In mid-September 2007, with a mean flux of $(97 \pm 15) \cdot 10^{-8}$ ph/cm$^2$/s (E $>$ 100 MeV), and a peak
flux of $(193 \pm 42) \cdot 10^{-8}$ ph/cm$^2$/s \cite{achen}. The second detection occurred in October 2007 \cite{giommi}, with a flux almost half than in the previous detection.
Rarely BL Lac objects undergo so large $\gamma$-ray activity. Considering the redhift of the source, the total power of the
jet is at a level comparable or higher than the maximum power extracted from a spinning black hole of $10^9$ solar masses \cite{vittorini}. The multiwavelenght campaigns show different
variability pattern for different bands: $\gamma$-ray variability was moderate during each observation. The hard X-ray flux remained almost unchanged. Instead \emph{Swift} observed
strong variability in soft X, and only moderate variability in the optical/UV. The correlation of optical and $\gamma$-ray variability suggests the emision
mechanism in the $\gamma$-ray being dominated by SSC. The overall variability pattern suggests that two SSC components are responsible for the non thermal emission \cite{vittorini}.
\section{Mkn 421} 
Mrk 421, at z=0.031, belongs to the class of High Energy Peaked BL Lac objects. It has been detected in the TeV range several times.
A flare in Hard X-rays from the source was detected on 2008 June the 10$^{th}$, followed by the detection of $\gamma$-ray activity \cite{donnarumma421}.
SuperAGILE measured an Hard X ray flux at the level of 55 mCrab (peak flux), an order of magnitude higher than the typical flux in quiescence. Integrating the GRID data for 5 days, we obtained a $\gamma$-ray flux of
$(42^{+14}_{-12}) \cdot 10^{-8}$ ph/cm$^2$/s (with significance 4.5). The
campaign was supported by the GASP consortium and by the VERITAS and MAGIC
team in the TeV range. The $\gamma$-ray detection triggered a ToO for \emph{Swift} in soft X. 
The wide multifrequency coverage, extending over a 12-decade spectral range,
allowed us \cite{donnarumma421}  to study the spectral energy distribution and
the  correlated light curves togheter. The 
June 2008 $\gamma$-ray flare has been interpreted as a rapid acceleration of leptons in the jet, 
in the framework of SSC model:
the Synchrotron component extending from optical to hard X data, and the SSC component extending above 100 MeV and peaked at 0.1 GeV.
\section{W Comae}
W Comae was detected in flaring activity by VERITAS on 2008 June the 8$^{th}$. AGILE repointed to the field and detected the source 24 hours after the VERITAS trigger.
A paper on that multiwavelength campaign is in preparation \cite{epian}.
\section{Conclusions}
AGILE succesfully detects blazars in flare. For some of them, i.e. 3C 454.3, PKS 1510-089, the emission features can be modelled in the framework of leptonic model,
including the External Compton component with seeds photons originated by the disk, but reprocessed by the Broad Line Regions. 
In particular, for 3C 454.3, we observed in December 2007 a so high $\gamma$-ray flux, that an inverse Compton component with seed photons coming from the hot corona is needed to explain the data.
In some other cases (for S5 0716+714, Mrk 421) the correlation of the variability patterns in optical and $\gamma$-ray, as well as the SED, suggest that the high energy peak of the Spectral Energy Distribution
is due to the SSC mechanism.
%
%
%

\end{document}